\documentclass{article}
\usepackage{graphicx}
\usepackage{bm}
\usepackage{amsfonts}
\usepackage{amsmath}
\usepackage{amsmath}
\usepackage{amssymb}

\begin{document}

\title{Relativistic MOND theory based on\\ the Khronon scalar field}

% for a single author use the following line
% \author{Speaker's Name\\ {\small  Institution's Name, Address} \\{\small {\it Email} : Speaker's E-mail}}

\author{Luc \textsc{Blanchet} and Sylvain \textsc{Marsat}\\
{\small
Institut d'Astrophysique de Paris --- UMR 7095 du CNRS,} 
\\{\small Universit\'e Pierre \& Marie Curie, 98\textsuperscript{bis} boulevard Arago, 75014 Paris, 
France} \\
{\small {\it Email} : blanchet@iap.fr, marsat@iap.fr}
}

%\date{\today}
\maketitle

\begin{abstract}
We investigate a model of modified gravity recovering the modified Newtonian dynamics (MOND) in the non-relativistic limit, based on the introduction of a preferred time foliation violating Lorentz invariance in the weak-field regime. Lorentz-invariance violation has been studied in the framework of Einstein-{\ae}ther theory, the generalization of which, known as non-canonical Einstein-{\ae}ther theory, having been proposed as a relativistic formulation of MOND. Our model can be seen as a minimal specialization to the hypersurface orthogonal case, which allows a different interpretation in terms of the preferred time : it can be either treated as a dynamical scalar field in a 4D formulation, or chosen as the time coordinate in a 3+1 formulation. We discuss the equivalence of the two points of view and the non-relativistic limit of the model.  
\end{abstract} 

\section{Motivation} 

As an alternative to the standard $\Lambda$-CDM model, performing well in cosmology and on large scales but leaving open questions at galactic scales, the Modified Newtonian Dynamics (MOND) \cite{Milg123} intends to answer the missing mass problem by a phenomenological modification of gravity in the weak-field, large distances regime. It has proven successfull concerning galactic observations, while needing some extension at the scale of clusters of galaxies and for cosmology (for reviews, see \cite{SandMcG02,FamMcG11}). We will adopt for MOND its modified Poisson-equation version \cite{BekM84}:
\begin{equation}\label{ModPoisson}
	\bm{\nabla}\cdot\left[ \mu\left( \frac{|\bm{g}|}{a_0}\right) \bm{\nabla}\phi \right]=4\pi G \rho \; ,
\end{equation}
where $\rho$ is the ordinary matter density, $\phi$ is the Newtonian potential, $\bm{g}=\bm{\nabla}\phi$ is the gravitational field, $a_0=1.2\,10^{-10}\,\mathrm{m\,s}^{-2}$ is a universal constant acceleration scale, and $\mu$ is an interpolation fonction of the ratio $x=|\bm{g}|/a_0$, with asymptotic behaviour : $\mu(x)\rightarrow1\;[x\rightarrow+\infty]$ and $\mu(x)\sim x\;[x\rightarrow 0]$.

This modified Poisson equation is a purely non-relativistic formula, while a relativistic theory is needed to address issues concerning cosmology and gravitational lensing. Several relativistic extensions for MOND have been proposed in the past, including a tensor-vector-scalar theory (TeVeS) \cite{Bek04,Sand97}, a bimetric theory \cite{Milgbim2}, non-canonical Einstein-\AE{}ther theories \cite{ZFS07,HZL08}, and a modified dark matter theory \cite{BL09}. 

With a completely different motivation, the Ho\v{r}ava-Lifshitz approach \cite{Horava09} introduced a Lorentz invariance breaking preferred foliation of spacetime to build a power-counting renormalizable theory of gravity in the strong-field regime. This approach has been improved and extended in \cite{BPS11}.

In this contribution, which is a summary of our recent paper \cite{BM11} (see also \cite{Sand11}) to which we refer for more details, we investigate a specific relativistic extension of MOND, that can be either understood as a Lorentz-invariance breaking theory using a preferred time foliation, or as a hypersurface-orthogonal restriction of non-canonical Einstein-\AE{}ther theories.  

\section{Non-canonical Einstein-\AE{}ther theories}

Einstein-\AE ther theories have been introduced as a phenomenological approach to Lorentz-invariance violation \cite{JM00}. A preferred direction of time is described by a dynamical, timelike unit vector $n^{\mu}$. The corresponding Lagrangian density is usually written as:\footnote{We use the $(-+++)$ convention for the metric, and often set $G=c=1$.}
\begin{equation}
	\mathcal{L}_{\AE}=\frac{\sqrt{-g}}{16\pi} \Bigl[ R + \mathcal{K} + \lambda(n^{\mu}n_{\mu} + 1) \Bigr] \; ,
\end{equation}
where $R$ is the Ricci scalar, $\lambda$ is a Lagrange multiplier enforcing the normalization condition $n^{\mu}n_{\mu} = -1$, and where $\mathcal{K}$ represents the most general Lagrangian density that is quadratic in the derivatives of $n^{\mu}$ :
\begin{subequations}\label{E-AEAction}
\begin{align}
	\mathcal{K} &= \mathcal{K}^{\mu \nu \rho \sigma} \nabla_{\mu}n_{\rho} \nabla_{\nu}n_{\sigma} \; ,\\
	\mathcal{K}^{\mu \nu \rho \sigma} &= c_1g^{\mu\nu}g^{\rho\sigma} + c_2 g^{\mu\rho}g^{\nu\sigma} + c_3 g^{\mu\sigma}g^{\nu\rho} + c_4 n^{\mu}n^{\nu}g^{\rho\sigma}\; ,
\end{align}
\end{subequations}
where $c_1$, $c_2$, $c_3$ and $c_4$ are dimensionless constants, left unspecified at this stage. Constraints in the Solar System computed from Parametrized Post-Newtonian parameters have been obtained in \cite{FJ06}.

The non-canonical generalization of the previous action has been proposed in \cite{ZFS07,HZL08}, as a relativistic extension of MOND. It consists of replacing the \ae{}ther action $\mathcal{K}$ by a free function $F(\mathcal{K})$, \text{i.e.} of considering the Lagrangian
\begin{equation}
	\mathcal{L}_{\text{non-canonical}\,\AE}=\frac{\sqrt{-g}}{16\pi} \Bigl[ R + F(\mathcal{K}) + \lambda(n^{\mu}n_{\mu} + 1) \Bigr] \; .
\end{equation}
More generally, one could imagine introducing several arbitrary functions $F_1$, $F_2$, $F_3$ and $F_4$ corresponding to the various terms in \eqref{E-AEAction}. In the non-relativistic limit, specifying the behaviour of the fonction $F(\mathcal{K})$ around $\mathcal{K}=0$ allows to recover the modified Poisson equation \eqref{ModPoisson}. The class of cosmologies that are possibly produced by this type of models has been extensively studied in \cite{Zuntz10}. 

\section{From the \AE{}ther to the Khronon}

Instead of directly using a timelike dynamical vector field $n^{\mu}$ in the action, Lorentz invariance violation can be realized by introducing a preferred foliation of spacetime. The link between these two formulations has been studied in \cite{Jacobson10}. The foliation consists of hypersurfaces of constant value of a scalar field $\tau$, which we will call the ``Khronon'' field following \cite{BPS11}. It defines a hypersurface-orthogonal unit vector field $n_{\mu}$, whose relation to the Khronon field $\tau$ is:
\begin{equation}\label{khronon}
	n_{\mu} = - N \,\partial_{\mu}\tau \; , ~~\text{with}~ \; N=\frac{1}{\sqrt{-g^{\rho\sigma}\partial_{\rho}\tau\partial_{\sigma}\tau}} \; .
\end{equation}
We see that the condition for the Khronon field to define a well-behaved spacetime foliation is that its gradient must remain timelike everywhere. Recall that, in general, a unit timelike \ae{}ther vector does not define a foliation of spacetime; the condition for it to be hypersurface-orthogonal is $n_{[\mu}\nabla_{\nu}n_{\rho]}=0$ (the so-called Frobenius theorem). Notice also that the fundamental ingredient of the theory is now the scalar field $\tau$, while in Einstein-\AE{}ther theory it is the \ae{}ther vector field, with three degrees of freedom (one being suppressed by the unit-norm constraint) instead of one. An advantage of the formulation \eqref{khronon} is that the normalization condition is automatically satisfied and we do not need a Lagrange multiplier in the action.

Given this spacetime foliation, further geometrical definitions are: 
\begin{equation}
	\gamma_{\mu}^{\phantom{\mu}\nu} = \delta_{\mu}^{\phantom{\mu}\nu} + n_{\mu}n^{\nu} \; , \quad K_{\mu\nu}=\gamma_{\mu}^{\phantom{\mu}\rho}\nabla_{\rho}n_{\nu} \;, \quad a_{\mu}=n^{\nu}\nabla_{\nu}n_{\mu} \;,
\end{equation}
where $\gamma_{\mu}^{\phantom{\mu}\nu}$ is the projector orthogonal to $n^{\mu}$, $K_{\mu\nu}$ is the hypersurface's extrinsic curvature tensor and $a_{\mu}$ is the spacelike four-acceleration of the congruence orthogonal to the hypersurface with velocity $n_{\mu}$. We also define the projected covariant derivative operator as, for instance:
%(for instance acting on a vector, with a straightforward generalization to tensors with a different rank): 
$D_{\mu}V^{\nu}=\gamma_{\mu}^{\phantom{\mu}\rho}\gamma_{\sigma}^{\phantom{\sigma}\nu}\nabla_{\rho}V^{\sigma}$. Those definitions could be done formally in the non-hypersurface-orthogonal case, but the following important relation is specific to this case:
\begin{equation}
	a_{\mu} = D_{\mu} \ln N
\end{equation}
and, in addition, the extrinsic curvature is symmetric: $K_{\mu\nu}=K_{\nu\mu}$.

We are now able to rewrite the four terms written in Eq.~\eqref{E-AEAction} as the basic ingredients of the Lagrangian density of the Khronon field, expressed in terms of extrinsic curvature and acceleration:
\begin{subequations}
\begin{align}
	\text{Term $c_1$ :} & ~\; \nabla_{\mu}n_{\nu}\nabla^{\mu}n^{\nu} = K_{\mu\nu}K^{\mu\nu}-a^2 \; ,\\
	\text{Term $c_2$ :} & ~\; (\nabla_{\mu}n^{\mu})^2 = K^2 \; ,\\
	\text{Term $c_3$ :} & ~\; \nabla_{\mu}n_{\nu}\nabla^{\nu}n^{\mu} = K_{\mu\nu}K^{\mu\nu} \; ,\\
	\text{Term $c_4$ :} & ~\; n^{\mu}n^{\nu}\nabla_{\mu}n_{\rho}\nabla_{\nu}n^{\rho} = a^2 \; ,
\end{align}
\end{subequations} 
where $a^2=a_{\mu}a^{\mu}$ and $K=K^{\phantom{\mu}\mu}_\mu$ is the trace of the extrinsic curvature. We see, as was pointed out in \cite{Jacobson10}, that only three out of these four terms are actually independent in the hypersurface-orthogonal case. We can go further by investigating the fates of the extrinsic curvature and acceleration in two regimes of interest, the non-relativistic or post-Newtonian limit and the homogeneous and isotropic cosmology. In the post-Newtonian limit and in adapted coordinates, \textit{i.e.} when the time coordinate $t$ is identified with $\tau$, one obtains (see below):
 \begin{equation}
	a_{i} = \frac{1}{c^2}\partial_i \phi + \mathcal{O}(4) \; , \quad\text{and}\quad K_{ij}, \; K = \mathcal{O}(3) \; ,
\end{equation} 
with $\phi$ the Newtonian potential and where $ \mathcal{O}(n)$ means terms of order at least $1/c^n$. On the other hand, when considering a perfectly homogeneous and isotropic (FLRW) Universe, where the foliation is identical to the cosmic time foliation, we have: 
 \begin{equation}
	a_{\mu} = 0 \; , \quad\text{and}\quad K_{ij}K^{ij}=3H^2, \; K^2 = 9H^2 \; ,
\end{equation}
where $H$ is the standard Hubble parameter.
 
This means that the $c_1$ and $c_4$ terms are the only ones to contribute in the non-relativistic regime, and that the $c_4$ term vanishes in the usual cosmological background. However, notice that the $c_1$ term switches sign between the two regimes, whereas the $c_4$ term is positive in any case. In the model discussed below, we use only this $a^2$ term as the basic ingredient to recover MOND in the non-relativistic limit as an alternative to dark matter at the galactic scale, leaving the cosmology aside.

\section{Specific example of MOND theory}

\subsection{Covariant formulation}

In the 4D covariant point of view, the model constitutes a modification of General Relativity (GR) by the introduction of an additional scalar field, the Khronon $\tau$. We introduce in the action a free function of the norm of the acceleration $a$:
\begin{equation}
	\mathcal{L} = \frac{\sqrt{-g}}{16\pi}\bigl[ R - 2f(a)\bigr] + \mathcal{L}_\text{m}[g_{\mu\nu},\Psi] \;.
\end{equation}
This choice corresponds to the $c_4$ term in the Einstein-\AE{}ther action. We also assume the standard coupling of matter fields $\Psi$ to the metric. The function $f(a)$ is to be specified later, in order to recover MOND in the weak-field limit and GR in the strong-field regime. Variation of the action with respect to the metric and to the $\tau$ field yields the field equations:
\begin{subequations}
\begin{align}
		G^{\mu\nu}+f(a)g^{\mu\nu} + 2n^{\mu}n^{\nu}\nabla_{\rho}\left[\chi(a) a^{\rho}\right] - 2\chi(a) a^{\mu}a^{\nu} = 8\pi \,T^{\mu\nu} \; ,\\
		\nabla_{\mu}\left[ n^{\mu}\nabla_{\nu}(\chi(a) a^{\nu})\right] = \frac{1}{2}n^{\nu}\nabla_{\nu}f + \chi(a) \,a^{\mu}a^{\nu}K_{\mu\nu} \; ,
\end{align}
\end{subequations}
with $T^{\mu\nu}$ the matter stress-energy tensor, and where we used the short-hand notation $\chi(a)\equiv f'(a)/2a$. The second equation, which we call the $\tau$-equation, is of fourth order in the derivatives. We will see however that, in adapted coordinates, it becomes of first order only in time derivatives of geometrical quantities. If, from the first equation, which we call the modified Einstein equation, we define an equivalent stress-energy tensor $T^{\mu\nu}_{\tau}$ for the Khronon field, one can see that the $\tau$-equation is in fact equivalent to $\nabla_{\nu}T^{\mu\nu}_{\tau}=0$. Hence, because of the Bianchi identity, the matter conservation equation $\nabla_{\nu}T^{\mu\nu}=0$ and the modified Einstein equation together contain the $\tau$-equation. 

\subsection{3+1 formulation}

We now write the 3+1 formulation of the theory in adapted coordinates, choosing $t=\tau$. With standard definitions for the lapse $N$, the shift $N_i$ and the spatial metric $\gamma_{ij}$, the 3+1 parametrization of the metric reads
\begin{equation}
	ds^2 = -(N^2-N_{i}N^{i})\mathrm{d} t^2 + 2N_{i}\mathrm{d} t \mathrm{d} x^{i} + \gamma_{ij}\mathrm{d} x^{i}\mathrm{d} x^{j} \; .
\end{equation}
In these adapted coordinates, we have $n_{\mu}=(-N,0)$ and $a_i = \partial_i \ln N$, with $N$ being now a geometrical quantity. The 3+1 form of the action is then:
\begin{equation} 
	\mathcal{L} = \frac{\sqrt{\gamma}}{16\pi}N\bigl[ \mathcal{R}+K_{ij}K^{ij}-K^{2}-2f(a)\bigr] + \mathcal{L}_\text{m}[N,N_{i},\gamma_{ij},\Psi] \;.
\end{equation}
The $\tau$ field is not a dynamical field anymore --- it has been absorbed into the time coordinate. Its contribution to the action is purely geometric, $f(a)$ depending now only on the gradient of the lapse $N$. This term explicitly breaks Lorentz invariance. Varying the action with respect to $N$, $N_i$ and $\gamma_{ij}$, we obtain:
\begin{subequations}\label{3+1FieldEq}
\begin{align}
	\mathcal{R} + K^2 - K_{ij}K^{ij} - 2f + 4\chi a^2 + 4D_{i}(\chi a^{i}) &= 16\pi \,\varepsilon \;,\\
	D_{j}\left( K^{ij}-\gamma^{ij}K \right) &=  - 8\pi J^{i} \;,\\
	\mathcal{G}^{ij} + \frac{1}{N}D_t\left(K^{ij}-\gamma^{ij}K\right)+ \frac{2}{N}D_{k}\left[N^{(i}(K^{j)k}-\gamma^{j)k}K)\right] \nonumber \\
	 + 2 K^{ik}K^{j}_{\phantom{j}k}-K K^{ij}- \frac{1}{2}\gamma^{ij}\left(K^{kl}K_{kl}+K^2\right) \nonumber \\
	 - \frac{1}{N} \left(D^iD^jN - \gamma^{ij} D_kD^k N\right) - 2\chi a^ia^j + f \gamma^{ij} &=  8\pi \mathcal{T}^{ij}\;,
\end{align}
\end{subequations}
where $\delta\mathcal{L}_{m}/\delta N\equiv -\sqrt{\gamma}\varepsilon$, $\delta\mathcal{L}_{m}/\delta N_{i}\equiv\sqrt{\gamma}J^{i}$ and $\delta\mathcal{L}_{m}/\delta \gamma_{ij}\equiv N\sqrt{\gamma}\mathcal{T}^{ij}/2$.

\subsection{Equivalence between the formulations}

The two formulations of our model seem quite different, since the $\tau$ field is a dynamical field in the 4D point of view while it becomes a mere time coordinate in the 3+1 point of view. It has been pointed out in \cite{Jacobson10} that the fact that the $\tau$-equation is contained in the modified Einstein equation and the conservation of the matter stress-energy tensor allows one to consistently treat it as a coordinate.

We explicitly checked this equivalence between the two formulations at the level of the field equations, by verifying that the 3+1 projection (after variation) of the 4D equations was in agreement with the 3+1 ones, obtained by the variation with respect to geometrical quantities after projection. We also checked that the 3+1 projection of the $\tau$-equation, namely 
\begin{equation}\label{Teq3D}
D_t\biggl[D_{i}(\chi a^{i}) + \chi a^2 - \frac{f}{2}\biggr] + N K \Bigl(D_{i}(\chi a^{i}) + \chi a^2\Bigr) - N \chi \,a^{i}a^{j}K_{ij} = 0 \; ,
\end{equation}
where $D_t\equiv\partial_t-N^kD_k$, was indeed contained in the field equations \eqref{3+1FieldEq} and the 3+1 writing of the conservation of matter stress-energy tensor. Notice that this equation \eqref{Teq3D} has now become of first order only in time derivatives of geometrical quantities.

\subsection{Recovering MOND in the non-relativistic regime}

For a system at rest with respect to the preferred frame, we may write, in the non-relativistic (NR) limit, restoring the $c$ factors:
\begin{equation}
		N = 1 + \frac{\phi}{c^2} + \mathcal{O}\left( 4 \right), \; ~N_{i} = \mathcal{O}\left( 3 \right), \; ~\gamma_{ij} = \delta_{ij}\left(1 - \frac{2\psi}{c^2}\right) + \mathcal{O}\left( 4 \right).
\end{equation}
Here $\phi$ and $\psi$ are the usual Newtonian potentials, and we recall that $a_{i} = \partial_i \phi /c^2 + \mathcal{O}(4)$. Combining the 3+1 field equations \eqref{3+1FieldEq}, we first obtain the equality of the Newtonian potentials, $\phi=\psi+\mathcal{O}(2)$, crucial for the dark matter seen by gravitational lensing \cite{SandMcG02}, and we get a modified Poisson-like equation,
\begin{equation}\label{modPoisson}
	D_{i}\left[(1+\chi)a^i\right] + f + a^2 - \frac{1}{N}D_t K - K^{ij}K_{ij} = 4\pi\left(\varepsilon + \frac{2}{N}N_iJ^i + \mathcal{T}\right)\; ,
\end{equation}
which reduces in the NR limit to $\bm{\nabla} \cdot \bigl[ \left( 1+ \chi \right) \bm{\nabla} \phi\bigr] = 4\pi G \rho + \mathcal{O}\left( 2 \right)$, with $\rho$ the ordinary rest-mass density of matter. We see that there is a one-to-one correspondence between the MOND $\mu$ function and the function $f$ in the action, namely $\mu=1+\chi$ where $\chi(a)=f'(a)/2a$. Requirements on $f$ to recover MOND are as follows ($\Lambda_{\infty}$ and $\Lambda_0$ being two constants):
\begin{subequations}
\begin{align}
f(a) &\sim \Lambda_{\infty} \quad \text{in the strong-field regime $a\rightarrow\infty$,} \label{fStrongField}\\
f(a) &= \Lambda_0 - a^2 + \frac{2 a^3}{3 a_0} + \mathcal{O}\left(\frac{a^4}{a_0^2}\right) \quad \text{in the weak-field regime $a\rightarrow 0$.}
\end{align}
\end{subequations}
Note that the usual strong-field condition would be $f'(a)/a \rightarrow 0$ (so that we recover the standard Poisson equation), but here we imposed a stronger condition when $a\rightarrow\infty$ in order to recover exactly GR with a cosmological constant $\Lambda_{\infty}$ in the strong-field regime. 
  
\subsection{Solar System effects}

We can get a rough estimate of the smallness of MOND effects in the Solar System, assuming that \eqref{fStrongField} holds, which as we said is in fact more restrictive than the usual MOND requirement $\chi(a)\rightarrow 0$. To this end we write (with $c=1$)
\begin{equation}\label{fStrongField1}
f(a)\simeq \Lambda_{\infty}+k a_{0}^{2} \left(\frac{a_0}{a}\right)^\alpha~~\text{(when $a\rightarrow\infty$)}\;,
\end{equation}
with $k$ a dimensionless number of order one. This constant is fixed in order of magnitude by the zero-point of the function $f$ and the numerical coincidence $\Lambda \sim a_{0}^{2}$. Then, translating the MOND equation in spherical symmetry as $(1+\chi)g=g_\text{N}$, with $g_\text{N}$ the Newtonian acceleration, and inverting, one obtains 
\begin{equation}
g\simeq g_\text{N}\biggl[1+ \frac{k\alpha}{2}\left(\frac{a_0}{g_\text{N}}\right)^{2+\alpha}\biggr]\;.
\end{equation}
Defining the MOND transition radius $R_0$ by $a_0\equiv GM/R_{0}^{2}$, we have that $(a_0/g_\text{N})^{2+\alpha}=(r/R_0)^{4+2\alpha}$. Since numerically $R_0\simeq 7100$ AU, we see that within a Neptune orbit ($\simeq 30$ AU), the relative effect is at most of the order of $10^{-12}$ for $\alpha$ close to $0$, and is at most $10^{-15}$ for a typical value $\alpha\simeq 1$. Relativistic corrections in the Solar System are of the order of $(v/c)^2\simeq 10^{-8}$, and typical constraints on the values of PPN parameters \cite{Will} correspond to deviations from GR of the order of $10^{-8}\times 10^{-4}=10^{-12}$. Our requirement \eqref{fStrongField1} for recovering GR implies that the theory passes the Solar-System tests.

\section{Conclusion}

We investigated a simple model for a relativistic extension of MOND based on a preferred time foliation. We discussed how it relates to the more general framework of non-canonical Einstein-\AE{}ther theories and why it selects out of it the minimal ingredients required for recovering MOND in the non-relativistic limit. We analyzed the two points of view for the model, 4D or 3D, and discussed their equivalence. We investigated the non-relativistic limit and estimated the size of Solar-System effects. Further work is needed to study the cosmology of the model, and possibly to extend it in order to account for the missing mass problem at large cosmological scales. 

%\bibliographystyle{plain}
%\bibliography{/Users/marsat/Documents/publications/bibliographie/ListeRef}

\end{document}